\documentclass{article}
\usepackage{amssymb}

\usepackage{amsmath}
\usepackage{graphicx}

\newtheorem{theorem}{Theorem}
\newtheorem{acknowledgement}[theorem]{Acknowledgement}

\input{tcilatex}

\begin{document}

\title{{\Large Hamiltonization of theories with degenerate coordinates.}}
\author{D.M. Gitman\thanks{%
Institute of Physics, University of Sao Paulo, Brazil; e-mail:
gitman@fma.if.usp.br} and\ I.V. Tyutin\thanks{%
Lebedev Physics Institute, Moscow, Russia; e-mail: tyutin@lpi.ru}}
\date{\today  }
\maketitle

\begin{abstract}
We consider a class of Lagrangian theories where part of the coordinates
does not have any time derivatives in the Lagrange function (we call such
coordinates degenerate). We advocate that it is reasonable to reconsider the
conventional definition of singularity based on the usual Hessian and,
moreover, to simplify the conventional Hamiltonization procedure. In
particular, in such a procedure, it is not necessary to complete the
degenerate coordinates with the corresponding conjugate momenta.
\end{abstract}

\section{Introduction}

The Hamiltonization of Lagrangian theories is an important preliminary step
towards their canonical quantization \cite{Dirac64,GitTy90,HenTe92}. The
procedure is quite different for nonsingular and singular theories. Whereas
for nonsingular theories such a procedure is, in fact, the well-known
Legendre transformation, the Hamiltonization of singular theories is
sometimes a more difficult task. The singularity property of a theory is
usually defined by the corresponding Hessian, which is zero in the singular
case. The Hamiltonization procedure also depends essentially on theory
structure. In particular, it depends on the highest orders of time
derivatives in the Lagrange function. In principle, the Hamiltonization
procedure is quite well developed for theories of arbitrary orders$\;N\geq1$
of time derivatives \cite{GitTyL85}. However, as we are going to
demonstrate, for a class of theories where part of the coordinates does not
have any time derivatives in the Lagrange function (we call such coordinates
degenerate) it is reasonable to reconsider the conventional definition of
singularity and, moreover, to simplify the conventional Hamiltonization
procedure. In particular, in such a procedure (we call it the generalized
Hamiltonization procedure) we do not complete the degenerate coordinates
with the corresponding conjugate momenta. Indeed, it seems exaggerate to
introduce a momentum for the variable $p$ in a theory whose Lagrange
function is $L=p\dot{q}-V\left( q,p\right) $ (the corresponding action
already has Hamiltonian form) and then to struggle with irrelevant
constraints, see relevant remarks in \cite{FadJa88}. We show that the
degenerate coordinates may be treated on the same footing as usual
velocities (or highest order time derivatives in the Lagrangian function).
In fact, some observations about the possibility of a special treatment of
the degenerate coordinates were already implicitly presented in literature.
In this regard one can recall that sometimes, in the course of the
Hamiltonization of the Maxwell theory, $A_{0}$ is considered a Lagrange
multiplier to a constraint and no conjugate momentum to $A_{0}$\ is
introduced, see for example \cite{Dirac64}. For theories with degenerate
coordinates, the generalized Hamiltonization procedure contains less stages
than the usual Hamiltonization procedure and needs less suppositions about
the theory structure. There exist some models to which only the generalized
Hamiltonization procedure is applicable. In connection with this, one ought
to say that almost all modern physical gauge models are theories with
degenerate coordinates.

The present paper is organized as follows: in Sect.II, using a simple but
instructive example, we advocate a new definition of singularity and
consider the possibility of simplifying the Hamiltonization for theories
with degenerate coordinates. Then in Sect.III we formulate the generalized
Hamiltonization procedure and new criteria for singularity in the general
case of Lagrangian theory with degenerate coordinates. In Sect.IV we
consider some relevant examples. In the Appendix we discuss an useful notion
and properties of auxiliary variables, the latter being often used in the
main text.

\section{Theories with degenerate coordinates}

\subsection{ Which theories are conventionally called singular?}

We first recall the conventional definition of singularity of a theory
through the example of a theory without higher order time derivatives for
which the action reads $S=\int Ldt,$ and the Lagrange function has the form $%
L=L\left( q,\dot{q}\right) $, where $q=(q^{a};a=1,2,...,n)$ is the set of
generalized coordinates and $\dot{q}=(\dot{q}^{a}\equiv dq^{a}/dt)$. In such
a case, the Hessian $M$ is used for the classification. Namely:

\begin{equation}
M=\det\left| \left| \frac{\partial^{2}L}{\partial\dot{q}^{a}\partial\dot {q}%
^{b}}\right| \right| =\QATOPD{\{}{.}{\neq0,\;\mathrm{nonsingular\;theory}}{%
=0,\;\mathrm{singular\;theory}}.   \label{1a}
\end{equation}
Whenever a theory is nonsingular according to the above definition, the
corresponding Euler-Lagrange equations of motion (EM) can be solved with
respect to the highest time derivatives (here with respect to second-order
derivatives) of all the coordinates. Indeed,

\begin{align}
& \frac{\delta S}{\delta q^{a}}=\frac{\partial L}{\partial q^{a}}-\frac{d}{dt%
}\frac{\partial L}{\partial\dot{q}^{a}}=0\Longrightarrow M_{ab}\ddot{q}%
^{b}\,=K_{a}\,=\frac{\partial L}{\partial q^{a}}-\frac{\partial ^{2}L}{%
\partial\dot{q}^{a}\partial q^{b}}\dot{q}^{b}\,;  \notag \\
& \mathrm{thus,\;}\frac{\delta S}{\delta q^{a}}\Longrightarrow\ddot{q}%
^{a}\,=M^{ab}K_{b}\,,\;M_{ab}=\frac{\partial^{2}L}{\partial\dot{q}%
^{a}\partial\dot{q}^{b}}\,,\;M^{ab}M_{bc}=\delta_{c}^{a}\,,   \label{2a}
\end{align}
which in turn means that the EM of a nonsingular theory of the above type
always have a unique solution whenever $2n$ initial data are given. The
Hamiltonization of nonsingular theories leads to the Hamilton EM without any
constraints on the phase-space variables $q,p$.

In the general case (theories with higher derivatives), the action reads:

\begin{align}
& S=\int Ldt,\;L=L\left( t,q^{\left( l\right) }\right) \,,  \notag \\
& q^{\left( l\right) }=\left( q^{a\left( l_{a}\right) }\equiv
d^{l_{a}}q^{a}/dt^{l_{a}}\right) ,\;a=1,...,n,\;l_{a}=0,1,...,\bar{N}_{a}\,. 
\label{5b}
\end{align}
Here $L$ depends on the coordinates $q^{a}=q^{a\left( 0\right) }$ and their
time derivatives $q^{a\left( l_{a}\right) }$ up to some finite orders $\bar{N%
}_{a}\,$. Bearing in mind the same Lagrange function (\ref{5b}), it is
sometimes convenient to assume it to be a function of the coordinates and
their time derivatives up to some finite order $N_{a}\geq\bar{N}_{a}\,,$
where $\bar{N}_{a}$ are the above mentioned orders of the time derivatives
that actually enter $L\,$. Thus, we introduce a set of theories with the
same Lagrange function $L$ but with different orders $\{N_{a}\}\,.$ From the
point of view of the Lagrangian formulation it is obvious that all theories
with the same $L$ and different orders $\{N_{a}\}$ are equivalent. Even
though their Hamiltonization involves different extended phase spaces, we
end up with equivalent formulations \cite{GitTyL85,GitTy90}.

A generalization of the definition (\ref{1a}) for theories with the Lagrange
function $L$ and orders $\{N_{a}\}$ was proposed in \cite{GitTyL85,GitTy90}.
Such a definition is based on a simple generalization of the Hessian,

\begin{equation}
M=\det\left| \left| \frac{\partial^{2}L}{\partial q^{a(N_{a})}\partial
q^{b(N_{b})}}\right| \right| =\QATOPD{\{}{.}{\neq0,\;\mathrm{%
nonsingular\;theory}}{=0,\;\mathrm{singular\;theory}}\,,\;N_{a}\geq1\,. 
\label{7c}
\end{equation}
When we effected the conventional Hamiltonization, we proceeded from a
system of first-order equations. To this end, we introduce new variables%
\begin{equation*}
x_{1}^{a}=q^{a\left( 0\right) }\,,\;x_{s}^{a},\,s=2,...,N_{a}\,;\;\upsilon
^{a}=q^{a\left( N_{a}\right) }\,, 
\end{equation*}
and impose the relations 
\begin{equation*}
\dot{x}_{s}^{a}=x_{s+1}^{a}\,,\;s=1,...,N_{a}-1\,;\;\dot{x}%
_{N_{a}}^{a}=\upsilon^{a}\,. 
\end{equation*}
The variables $\upsilon^{a}$ are called velocities. The variational
principle for the initial action $S=\int Ldt$ is equivalent to the one for
the first-order action $S^{\upsilon},$%
\begin{align}
& S^{\upsilon}=\int\left\{ L^{\upsilon}+\sum_{a}\left[
\sum_{s=1}^{N_{a}-1}p_{a}^{s}\left( \dot{x}_{s}^{a}-x_{s+1}^{a}\right)
+p_{a}^{N_{a}}\left( \dot{x}_{N_{a}}^{a}-\upsilon^{a}\right) \right]
\right\} dt  \notag \\
& =\int\left\{ \sum_{a}\sum_{s=1}^{N_{a}}p_{a}^{s}\dot{x}_{s}^{a}-H^{%
\upsilon}\right\} dt\,,  \notag \\
& L^{\upsilon}=\left. L\right| _{q^{a\left( s-1\right) }\rightarrow
x_{s}^{a},\;\,q^{a\left( N_{a}\right)
}\rightarrow\upsilon^{a}}\;,\,\;H^{\upsilon}=\sum_{a}\left[
\sum_{s=1}^{N_{a}-1\,}p_{a}^{s}x_{s+1}^{a}+p_{a}^{N_{a}}\upsilon^{a}\,\right]
-L^{\upsilon}\,.   \label{10c}
\end{align}
The momenta $\,p$ appear as Lagrange multipliers to the new imposed
equations.\ The pairs $x\,,\,p$ form the phase space and all the variables $%
x\,,\,p\,,\upsilon$ form the extended phase space. The corresponding
Euler-Lagrange EM read:

\begin{align}
\left. 
\begin{array}{c}
\frac{\delta S^{\upsilon}}{\delta p_{a}^{s}}=\dot{x}_{s}^{a}-x_{s+1}^{a}=0,%
\;s=1,...,N_{a}-1 \\ 
\frac{\delta S^{\upsilon}}{\delta p_{a}^{N_{a}}}=\dot{x}_{N_{a}}^{a}-%
\upsilon^{a}=0%
\end{array}
\right\} & \Longrightarrow\left\{ 
\begin{array}{c}
\dot{x}_{s}^{a}=\left\{ x_{s}^{a}\,,H^{\upsilon}\right\} \\ 
s=1,...,N_{a}%
\end{array}
\right. ;  \notag \\
\left. 
\begin{array}{c}
\frac{\delta S^{\upsilon}}{\delta x_{1}^{a}}=\frac{\partial L^{\upsilon}}{%
\partial x_{1}^{a}}-\dot{p}_{1}=0 \\ 
\frac{\delta S^{\upsilon}}{\delta x_{s}^{a}}=\frac{\partial L^{\upsilon}}{%
\partial x_{s}^{a}}-\,p_{a}^{s-1}-\dot{p}_{a}^{s}=0,\;s=2,...,N_{a}%
\end{array}
\right\} & \Longrightarrow\left\{ 
\begin{array}{c}
\dot{p}_{a}^{s}=\left\{ p_{a}^{s}\,,\,H^{\upsilon}\right\} \\ 
s=1,...,N_{a}%
\end{array}
\right. ;  \notag \\
\frac{\delta S^{\upsilon}}{\delta\upsilon^{a}}=\frac{\partial L^{\upsilon}}{%
\partial\upsilon^{a}}-p_{a}^{N_{a}}=0 & \Longrightarrow\frac{\partial
H^{\upsilon}}{\partial\upsilon^{a}}=0\,.   \label{11a}
\end{align}
The set of\ the variables$\;(x_{s}^{a},\;s=2,...,N_{a}\,,\;p_{a}^{s}\;,%
\;s=1,...,N_{a},\;\upsilon^{a})\;$is\ the\ auxiliary\ one\ (see Appendix)
and\ can be\ excluded\ from\ EM\ and\ the\ action\ so\ that\ we\ obtain\
the\ formulation\ based\ on\ the action$\;$(\ref{5b}).When performing the
Hamiltonization, we try to eliminate the velocities $\upsilon$\ from the set
(\ref{11a}). In nonsingular theories, according to the definition (\ref{7c}%
), it is possible to express all the velocities by means of the last set of
Eq. (\ref{11a}) as $\upsilon =\bar{\upsilon}\left( x,p_{a}^{N_{a}}\right) .$
In that case, all the velocities are auxiliary variables. They can be
excluded from the action (\ref{10c}). Thus, we arrive at the Hamilton action 
$S_{\mathrm{H}}$ and at the Hamilton EM for unconstrained phase-space
variables $x_{s}^{a},p_{a}^{s}$:

\begin{align*}
S_{\mathrm{H}} & =\int\left[ \sum_{a}\sum_{s=1}^{N_{a}}p_{a}^{s}\dot{x}%
_{s}^{a}-H\right] dt\,,\;H=\left. H^{\upsilon}\right| _{\upsilon =\bar{%
\upsilon}}\,, \\
\dot{x}_{s}^{a} & =\left\{ x_{s}^{a}\,,H\right\} \,,\,\,\,\dot{p}%
_{a}^{s}=\left\{ p_{a}^{s}\,,\,H\right\} \,.
\end{align*}
First, the Hamiltonization of nonsingular theories with higher-order time
derivatives was presented in \cite{Ostro50}. The Hamiltonization of singular
theories with higher-order time derivatives, on the base of the action (\ref%
{10c}), was considered in \cite{GitTyL85,GitTy90}.

\subsection{Degenerate coordinates. An instructive example}

Let us now suppose that some of the generalized coordinates do not have any
time derivatives in the Lagrange function. In the general case (\ref{5b}),
that means that $\bar{N}_{a}$ are zero for some of that coordinates. We
shall call the coordinates with $\bar{N}_{a}=0$ degenerate.\emph{\ }%
According to the conventional definitions (\ref{1a}) or (\ref{7c})$,$ any
theory with degenerate coordinates is singular. However, here we are going
to discuss the following question: is it always reasonable to treat theories
with degenerate coordinates as singular and to follow the above described
the conventional Hamiltonization procedure? To answer this question it is
instructive to first consider a class of theories with two coordinates $x,u$
and with Lagrange functions of the form

\begin{equation}
L=L\left( x,\dot{x},u\right) .   \label{34}
\end{equation}
Here the Hessian (\ref{1a}) is zero, $M=0$, therefore we are formally
dealing with the singular case. Nevertheless, we can demonstrate that the
corresponding Euler-Lagrange EM

\begin{align}
\frac{\delta S}{\delta x} & =\frac{\partial L}{\partial x}-\frac{\partial
^{2}L}{\partial\dot{x}\partial x}\dot{x}-\frac{\partial^{2}L}{\partial\dot {x%
}\partial u}\dot{u}-\frac{\partial^{2}L}{\partial\dot{x}\partial\dot{x}}%
\ddot{x}=0\,,  \label{5a} \\
\frac{\delta S}{\delta u} & =\frac{\partial L}{\partial u}=0   \label{6a}
\end{align}
have a unique solution (thus the theory is not a gauge theory) whenever the
determinant $\tilde{M}\mathcal{\,}$(we call it further the generalized
Hessian)

\begin{equation}
\tilde{M}=\det\left| \left| 
\begin{array}{cc}
\frac{\partial^{2}L}{\partial\dot{x}^{2}} & \frac{\partial^{2}L}{\partial 
\dot{x}\partial u} \\ 
\frac{\partial^{2}L}{\partial u\partial\dot{x}} & \frac{\partial^{2}L}{%
\partial u^{2}}%
\end{array}
\right| \right| \,=\frac{\partial^{2}L}{\partial\dot{x}^{2}}\frac{\partial
^{2}L}{\partial u^{2}}-\left( \frac{\partial^{2}L}{\partial\dot{x}\partial u}%
\right) ^{2}=\tilde{M}\left( x,\dot{x},u\right) \,,   \label{7}
\end{equation}
is not zero and two initial data are given. Indeed, the condition $\tilde
{M%
}\neq0$ necessarily implies either case (a) or case (b):

\begin{align}
& a)\;\frac{\partial^{2}L}{\partial u^{2}}\neq0\,,  \label{7a} \\
& b)\;\frac{\partial^{2}L}{\partial\dot{x}\partial u}\neq0\,.   \label{7b}
\end{align}

First consider case (a). In this case the equation (\ref{6a}) can be solved
with respect to $u$ ,

\begin{equation}
\frac{\partial L}{\partial u}=0\Longrightarrow u=\bar{u}\left( x,\dot {x}%
\right) \,,   \label{8a}
\end{equation}
and the equation

\begin{equation}
\frac{d}{dt}\frac{\partial L}{\partial u}=\frac{\partial^{2}L}{\partial
u\partial x}\dot{x}+\frac{\partial^{2}L}{\partial u\partial\dot{x}}\ddot {x}+%
\frac{\partial^{2}L}{\partial u^{2}}\dot{u}=0\,   \label{8c}
\end{equation}
can be solved with respect to $\dot{u},$

\begin{equation}
\dot{u}=-\left( \frac{\partial^{2}L}{\partial u^{2}}\right) ^{-1}\left[ 
\frac{\partial^{2}L}{\partial u\partial x}\dot{x}+\frac{\partial^{2}L}{%
\partial u\partial\dot{x}}\ddot{x}\right] \,.   \label{9a}
\end{equation}
Substituting (\ref{9a}) into (\ref{5a}), we arrive at the following equation

\begin{align*}
& \tilde{M}\left( x,\dot{x},u\right) \ddot{x}=F_{1}\left( x,\dot {x}%
,u\right) \,,\; \\
& F_{1}\left( x,\dot{x},u\right) =\frac{\partial L}{\partial x}\frac{%
\partial^{2}L}{\partial u^{2}}+\left[ \frac{\partial^{2}L}{\partial u\partial%
\dot{x}}\frac{\partial^{2}L}{\partial u\partial x}-\frac{\partial ^{2}L}{%
\partial x\partial\dot{x}}\frac{\partial^{2}L}{\partial u^{2}}\right] \dot{x}%
\,.
\end{align*}
Since $\tilde{M}\neq0$, the Euler-Lagrange EM can be reduced to the form

\begin{equation}
\ddot{x}=F_{1}\left( x,\dot{x},\bar{u}\right) /\tilde{M}\left( x,\dot {x},%
\bar{u}\right) \,,\;u=\bar{u}\left( x,\dot{x}\right) \,.   \label{10}
\end{equation}
They have a unique solution whenever two initial data are given, for
example, $x$ and $\dot{x}$ at the initial time instant.

Let us turn to the case (b). Here, due to (\ref{7b}), the equation (\ref{6a}%
) can be solved with respect to $\dot{x},$

\begin{equation}
\dot{x}=\bar{\upsilon}\left( x,u\right) \,,   \label{10a}
\end{equation}
and the equation

\begin{equation}
\frac{d}{dt}\frac{\partial L}{\partial u}=\frac{\partial^{2}L}{\partial
u\partial x}\dot{x}+\frac{\partial^{2}L}{\partial u\partial\dot{x}}\ddot {x}+%
\frac{\partial^{2}L}{\partial u^{2}}\dot{u}=0\,   \label{11}
\end{equation}
can be solved with respect to $\ddot{x}$ ,

\begin{equation}
\ddot{x}=-\left( \frac{\partial^{2}L}{\partial u\partial\dot{x}}\right) ^{-1}%
\left[ \frac{\partial^{2}L}{\partial u\partial x}\dot{x}+\frac{\partial ^{2}L%
}{\partial u^{2}}\dot{u}\right] \,.   \label{12}
\end{equation}
We may substitute (\ref{12}) into (5a) to get

\begin{align*}
& \tilde{M}\left( x,\dot{x},u\right) \dot{u}=F_{2}\left( x,\dot {x},u\right)
\,, \\
& F_{2}\left( x,\dot{x},u\right) =\frac{\partial L}{\partial x}\frac{%
\partial^{2}L}{\partial u\partial\dot{x}}+\left[ \frac{\partial^{2}L}{%
\partial x\partial\dot{x}}\frac{\partial^{2}L}{\partial u\partial\dot{x}}-%
\frac{\partial^{2}L}{\partial u\partial x}\frac{\partial^{2}L}{\partial \dot{%
x}^{2}}\right] \dot{x}\,.
\end{align*}
Since $\tilde{M}\neq0$, the Euler-Lagrange EM can be reduced to the form

\begin{equation}
\dot{u}=F_{2}\left( x,\bar{\upsilon},u\right) /\tilde{M}\left( x,\bar{%
\upsilon},u\right) \,,\;\dot{x}=\bar{\upsilon}\left( x,u\right) \, 
\label{13}
\end{equation}
They again have a unique solution whenever two initial data are given, for
instance, $x$ and $u$ at the initial time instant. One ought to remark that
provided both\ conditions\ a)\ and\ b) are satisfied, the\ EM can be written
in both forms (\ref{10}) and (\ref{13}).

Let us turn to the Hamiltonization of theories under consideration. First we
consider the conventional Hamiltonization procedure \cite%
{Dirac64,GitTy90,HenTe92}, that is, we choose $N_{x}=N_{u}=1$. In the
first-order formalism, the phase space is formed by the pairs $%
x,p;\,u,p^{\prime},$ and the extended phase space is formed by the variables 
$x,p;\,u,p^{\prime};\upsilon,\upsilon^{\prime}.$ The first-order formalism
action reads:

\begin{align}
& S^{\upsilon\upsilon^{\prime}}=\int\left[ L^{\upsilon\upsilon^{\prime}}+p%
\left( \dot{x}-\upsilon\right) +p^{\prime}\left( \dot{u}-\upsilon
^{\prime}\right) \right] dt=\int\left[ p\dot{x}+p^{\prime}\dot {u}%
-H^{\upsilon\upsilon^{\prime}}\right] dt\,,  \notag \\
& L^{\upsilon\upsilon^{\prime}}=L\left( x,\upsilon,u\right)
,\;\,\;H^{\upsilon\upsilon^{\prime}}=p\upsilon+p^{\prime}\upsilon^{\prime
}-L^{\upsilon\upsilon^{\prime}}\,.   \label{14}
\end{align}
When performing the Hamiltonization, we have to try to eliminate the
velocities $\upsilon,\upsilon^{\prime}$\ from the action $S^{\upsilon
\upsilon^{\prime}}$ and from the Euler-Lagrange EM

\begin{align}
& \frac{\delta S^{\upsilon\upsilon^{\prime}}}{\delta p}=0\Longrightarrow 
\dot{x}=\left\{ x,H^{\upsilon\upsilon^{\prime}}\right\} ,\;\;\frac{\delta
S^{\upsilon\upsilon^{\prime}}}{\delta x}=0\Longrightarrow\dot{p}=\left\{
p,H^{\upsilon\upsilon^{\prime}}\right\} \,,  \notag \\
& \frac{\delta S^{\upsilon\upsilon^{\prime}}}{\delta p^{\prime}}%
=0\Longrightarrow\dot{u}=\left\{ u,H^{\upsilon\upsilon^{\prime}}\right\}
,\;\;\frac{\delta S^{\upsilon\upsilon^{\prime}}}{\delta u}=0\Longrightarrow 
\dot{p}^{\prime}=\left\{ p^{\prime},H^{\upsilon\upsilon^{\prime}}\right\} \,,
\label{15} \\
& \frac{\delta S^{\upsilon\upsilon^{\prime}}}{\delta\upsilon}=-\frac{%
\partial H^{\upsilon\upsilon^{\prime}}}{\partial\upsilon}=\frac{\partial
L^{\upsilon \upsilon^{\prime}}}{\partial\upsilon}-p\,=0\,,\;\frac{\delta
S^{\upsilon \upsilon^{\prime}}}{\delta\upsilon^{\prime}}=-\frac{\partial
H^{\upsilon \upsilon^{\prime}}}{\partial\upsilon^{\prime}}=-p^{\prime}=0\,, 
\label{15a}
\end{align}
generated by the action $S^{\upsilon\upsilon^{\prime}}.$ The Hessian is
zero, the theory is singular and we cannot exclude both velocities $\upsilon
,\upsilon^{\prime}$ using Eqs. (\ref{15a}).

Let us suppose,\ however, that the generalized Hessian (\ref{7}) is not zero 
$(\tilde{M}\left( x,\upsilon,u\right) \neq0)$. Consider the following two
possible cases:

a)$\;\partial^{2}L^{\upsilon\upsilon^{\prime}}/\partial\upsilon^{2}=0.\;$%
Then $L^{\upsilon\upsilon^{\prime}}=\upsilon f_{1}\left( x,u\right)
-f_{2}\left( x,u\right) ,\;$and $\tilde{M}=-\left( \partial f_{1}/\partial
u\right) ^{2}\neq0\Longrightarrow\partial f_{1}/\partial u\neq0$.\
Therefore, the equation $\partial
L^{\upsilon\upsilon^{\prime}}/\partial\upsilon -p\,=0\Longrightarrow
f_{1}\left( x,u\right) -p=0$ can be solved with respect to $u$ as $u=\bar{u}%
\left( x,p\right) .$ Thus, we\ have\ two\ primary\ second-class constraints,

\begin{equation}
\Phi_{1}^{\left( 1\right) }=p^{\prime}=0,\;\;\Phi_{2}^{\left( 1\right) }=u-%
\bar{u}\left( x,p\right) =0,   \label{15b}
\end{equation}
and both velocities $\upsilon^{\prime},\upsilon$\ appear to be Lagrangian
multipliers in the total Hamiltonian $H^{\left( 1\right) }=f_{2}+\lambda
^{1}\Phi_{1}^{\left( 1\right) }$ $+\lambda^{2}\Phi_{2}^{\left( 1\right) }$
which defines now the Hamilton dynamics of the phase-space variables. No
more constraints appear. The constraints (\ref{15b}) have a special form %
\cite{GitTy90} and can be used to exclude variables $p^{\prime}$ and $u$
from the action and from the EM. Namely, we can substitute $p^{\prime}=0$\
and $u=\bar{u}\left( x,p\right) $ directly into $H^{\left( 1\right) }$ to
get the Hamiltonian $H=f_{2}\left( x,\bar{u}\right) ,$ which defines the
Hamilton dynamics of the remaining phase-space variables $x,p$ as: $\dot
{x}%
=\left\{ x,H\right\} \,,\;\dot{p}=\left\{ p,H\right\} \,.$

b)$\;\partial^{2}L^{\upsilon\upsilon^{\prime}}/\partial\upsilon^{2}\neq0$
(we should suppose that this condition holds in a vicinity of the\ point $%
x=\upsilon=u=0).\;$In\ this\ case\ the equation $\delta S^{\upsilon
\upsilon^{\prime}}/\delta\upsilon=\partial
L^{\upsilon\upsilon^{\prime}}/\partial\upsilon-p\,=0$ can be solved with
respect to $\upsilon$ as $\upsilon=\bar{\upsilon}\left( x,u,p\right) $ and
one primary constraint appears $\Phi^{\left( 1\right) }=p^{\prime}=0.$ The
total Hamiltonian that defines now the Hamilton dynamics of the phase-space
variables reads: $H^{\left( 1\right) }=p\bar{\upsilon}-L\left( x,\bar{%
\upsilon},u\right) +\lambda\Phi^{\left( 1\right) }.$ The consistency
condition for the primary constraint gives a secondary constraint $\left\{
p^{\prime},H^{\left( 1\right) }\right\} =\left. \partial
L^{\upsilon\upsilon^{\prime}}/\partial u\right| _{\upsilon=\bar{\upsilon}%
}=0\,,$ which can be solved with respect to $u$ as $\Phi^{\left( 2\right)
}=u-\bar{u}\left( x,p\right) =0.$ The variables $p^{\prime}$ and $u$ can be
excluded as in the previous case, and so we get a similar result. Thus,
after the conventional Hamiltonization we are left with the set of equations

\begin{align}
& \dot{x}=\left\{ x,H\right\} \,,\;\dot{p}=\left\{ p,H\right\} \,,\;u=\bar{u}%
\left( x,p\right) \,,  \notag \\
& \upsilon=\bar{\upsilon}\left( x,p\right) \,,\;H=p\bar{\upsilon}-L\left( x,%
\bar{\upsilon},\bar{u}\right) \,.   \label{17}
\end{align}

We see that whenever the determinant (\ref{7}) is not zero, the sector $x,p$
of the theory is not singular (no constraints on $x,p$), and the coordinate $%
u$ can be treated as an auxiliary variable. The number of initial data for
the EM is two. This fact matches the aforementioned Lagrangian treatment.

Moreover, in the present case, the conventional Hamiltonization procedure
can be simplified (we call new Hamiltonization procedure the generalized
one). Indeed, we may choose $N_{u}=\bar{N}_{u}=0$. Since the derivative $%
\dot{u}$ is not present in the Lagrange function, we do not introduce the
corresponding Lagrange multiplier and only introduce the $x$-velocity $%
\upsilon$. Then the equivalent first-order action reads:

\begin{align}
& S^{\upsilon}=\int\left[ L^{\upsilon}+p\left( \dot{x}-\upsilon\right) %
\right] dt=\int\left[ p\dot{x}-H^{\upsilon}\right] dt\,,  \notag \\
& L^{\upsilon}=L\left( x,\upsilon,u\right)
,\;\,\;H^{\upsilon}=p\upsilon-L^{\upsilon}\,.   \label{18ab}
\end{align}
Thus, the only conjugate momentum introduced is the $x$-momentum $p$. The
phase space is formed by $x,p$ and the extended phase space can be thought
of as $x,p;\upsilon,u.$ In the course of the Hamiltonization, it is natural
to treat both $\upsilon$ and $u$ on equal footing and try to exclude them
from the corresponding action and from the Euler-Lagrange EM

\begin{align}
& \frac{\delta S^{\upsilon}}{\delta p}=0\Longrightarrow\dot{x}=\left\{
x,H^{\upsilon}\right\} \,,\;\frac{\delta S^{\upsilon}}{\delta x}%
=0\Longrightarrow\dot{p}=\left\{ p,H^{\upsilon}\right\} \,,  \label{18} \\
& \frac{\delta S^{\upsilon}}{\delta\upsilon}=-\frac{\partial H^{\upsilon}}{%
\partial\upsilon}=\frac{\partial L^{\upsilon}}{\partial\upsilon }-p\,=0\,,\;%
\frac{\delta S^{\upsilon}}{\delta u}=-\frac{\partial H^{\upsilon}}{\partial u%
}=\frac{\partial L^{\upsilon}}{\partial u}=0\,.   \label{18a}
\end{align}

Consider the case $\tilde{M}\,\neq0.$ In this case$,$ the equations (\ref%
{18a}) may be used to express both $\upsilon$ and $u$ via the canonical pair 
$x,p$ as $\upsilon=\bar{\upsilon}\left( x,p\right) \,,\;u=\bar{u}\left(
x,p\right) $. We see that the variables $\upsilon,$ $u$ are auxiliary and
can be eliminated \ from the action $S^{\upsilon}$ (see the Appendix) to
obtain the action in the Hamiltonian form. The corresponding Hamiltonian $H$
\ is obtained by substituting $\upsilon=\bar{\upsilon}\left( x,p\right) \,,\
u=\bar{u}\left( x,p\right) $ directly into $H^{\upsilon}$ to get $H=p\bar{%
\upsilon}-L\left( x,\bar{\upsilon},\bar{u}\right) =H\left( x,p\right) \,\ $%
which determines the Hamilton dynamics of the pair $x,p$. Thus, after such
generalized Hamiltonization we arrive at the same set of equations (\ref{17}%
). Notice that in the generalized Hamiltonization procedure we did not use
the condition: $\partial^{2}L^{\upsilon}/\partial\upsilon ^{2}\neq0$ in a
vicinity of the\ point $x=\upsilon=u=0.$

Consider the case $\tilde{M}\,=0.$ Suppose, for example, that the rank of
the generalized Hessian matrix is zero and $\partial^{2}L^{\upsilon}/%
\partial
u^{2}=\partial^{2}L^{\upsilon}/\partial\upsilon^{2}=\partial^{2}L^{\upsilon
}/\partial\upsilon\partial u=0.$ In such a case the equations (\ref{18a})
appear to be constraints%
\begin{equation}
p-\frac{\partial L^{\upsilon}}{\partial\upsilon}\,=\Phi_{1}^{\left( 1\right)
}\left( x,p\right) =0\,,\;-\frac{\partial L^{\upsilon}}{\partial u}=\Phi
_{2}^{\left( 1\right) }\left( x\right) =0\,,   \label{19a}
\end{equation}
and the action (\ref{18ab}) corresponds to a Hamilton theory with primary
constraints (\ref{19a}),%
\begin{align}
& S^{\left( 1\right) }=\int\left[ p\dot{x}-H^{\left( 1\right) }\right]
dt\,,\;H^{\left( 1\right) }=H\left( x\right) +\lambda^{a}\Phi_{a}^{\left(
1\right) },\;a=1,2\,,  \notag \\
& H\left( x\right) =-L\left( x,0,0\right) ,\;\lambda^{1}=\upsilon
,\;\lambda^{2}=u\,.   \label{19}
\end{align}
Here the degenerate variable $u$ appears to be a Lagrange multiplier as well
as the velocity $\upsilon.$ If the rank of the generalized Hessian matrix is
one then only one primary constraint appears, and either $\upsilon$ or $u$
play the role of the Lagrange multipliers (the general case is considered
below). Further Hamiltonization is related to the usual Dirac procedure.

What can we learn from the above considerations? First of all, it is not
necessary to introduce the momenta associated to the degenerate coordinates
in the course of Hamiltonization. The procedure of the Hamiltonization can
be simplified. Moreover, the new generalized Hamiltonization procedure
motivates us to change the definition of singularity (\ref{1a}) in the
presence of degenerate coordinates. In this case, it is more reasonable to
classify theories according to the generalized Hessian (\ref{7}) and to
consider them nonsingular whenever $\tilde{M}\neq0$. Indeed, besides the
natural consistency with the generalized Hamiltonization procedure, the
generalized-Hessian criterion allows one to conclude immediately that a
theory is not a gauge theory whenever $\tilde{M}\neq0$. Looking upon the
conventional Hessian, we cannot come to such a conclusion without additional
analysis of the constraint structure. Below we present a generalization of
the conventional Hamiltonization procedure and a generalized singularity
criterion for a general Lagrangian theory.

\section{Generalized Hamiltonization procedure}

As was already said, in the general case the action reads $S=\int Ldt,$ the
Lagrange function has the form (\ref{5b}), and let the orders of the highest
derivatives be $\{N_{a}\}$. The corresponding Euler-Lagrange EM are

\begin{equation}
\frac{\delta S}{\delta q^{a}}=\sum_{l=0}^{N_{a}}\left( -1\right) ^{l}\frac{%
d^{l}}{dt^{l}}\left[ \frac{\partial L}{\partial q^{a\left( l\right) }}\right]
=0\,.   \label{20}
\end{equation}
We propose to classify Lagrangian theories as singular or nonsingular using
the generalized Hessian $\tilde{M}\,$:

\begin{equation}
\tilde{M}=\det\left| \left| \frac{\partial^{2}L}{\partial
q^{a(N_{a})}\partial q^{b(N_{b})}}\right| \right| =\QATOPD{\{}{.}{\neq0,\;%
\mathrm{nonsingular\;theory}}{=0,\;\mathrm{singular\;theory}}%
\,,\;N_{a}\geq0\,.   \label{21}
\end{equation}
We stress that the orders $N_{a}$ can be zero in the presence of degenerate
coordinates. The difference between definition (\ref{21}) and definition (%
\ref{7c})\ is related namely to the possibility of $N_{a}$ to be zero. If we
restrict all $N_{a}$ $\geq1$ even in the presence of degenerate
coordinates,\ the generalized Hessian (\ref{21}) and the Hessian (\ref{7c})
coincide. In what follows, we consider the Hamiltonization of general
theories according to the generalized Hamiltonization procedure and present
arguments in favour of the definition (\ref{21}). In particular, we will
demonstrate that in the nonsingular case (according to (\ref{21})) the
Euler-Lagrange EM always have a unique solution under an appropriate choice
of initial data and the Hamiltonization leads to usual Hamilton EM without
any constraints in the appropriate phase space.

Let us turn to the generalized Hamiltonization procedure. In the beginning$,$
we pass to the first-order formulation, which differs from the one
considered above whenever some of $N_{a}$ are zero. To this end, we divide
all the indices $a$, numbering the coordinates, into two groups,

\begin{equation}
a=\left( \bar{a},\underline{a}\right) \,,\;N_{\bar{a}}=0\,,\;N_{\underline {a%
}}\geq1\,,   \label{22}
\end{equation}
introduce new variables%
\begin{align*}
& x_{1}^{\underline{a}}=q^{\underline{a}\left( 0\right) }\,,\;x_{s}^{%
\underline{a}}\,,\;s=2,...,N_{\underline{a}}\,;\;\upsilon^{\underline{a}}=q^{%
\underline{a}\left( N_{\underline{a}}\right) }\,, \\
& \upsilon^{\bar{a}}=q^{\bar{a}\left( N_{\bar{a}}\right) }=q^{\bar{a}\left(
0\right) }=q^{\bar{a}}\,\;;\;\upsilon^{a}=\left( \upsilon^{\underline{a}%
},\upsilon^{\bar{a}}\right) \,,
\end{align*}
and impose the relations 
\begin{equation}
\dot{x}_{s}^{\underline{a}}=x_{s+1}^{\underline{a}}\,,\;s=1,...,N_{%
\underline {a}}-1\,;\;\dot{x}_{N_{\underline{a}}}^{\underline{a}}=\upsilon^{%
\underline {a}}\,.   \label{23}
\end{equation}
The variables $\upsilon^{a}$ will be called velocities. Thus, the degenerate
coordinates, for which we select $N=0,\;$have the status of velocities in
the first-order formulation. The variational principle for the initial
action $S\;$is equivalent to the one for the first-order action $%
S^{\upsilon},$%
\begin{align}
& S^{\upsilon}=\int\left[ L^{\upsilon}+\sum_{\underline{a}}\sum _{s=1}^{N_{%
\underline{a}}-1}p_{\underline{a}}^{s}\left( \dot{x}_{s}^{\underline{a}%
}-x_{s+1}^{\underline{a}}\right) +p_{\underline{a}}^{N_{\underline{a}%
}}\left( \dot{x}_{N_{\underline{a}}}^{\underline{a}}-\upsilon^{\underline{a}%
}\right) \right] dt  \notag \\
& =\int\left[ \sum_{\underline{a}}\sum_{s=1}^{N_{\underline{a}}}p_{%
\underline{a}}^{s}\dot{x}_{s}^{\underline{a}}-H^{\upsilon}\right] dt\,, 
\notag \\
& L^{\upsilon}=\left. L\right| _{q^{\underline{a}\left( s-1\right)
}\rightarrow x_{s}^{\underline{a}},\,q^{a\left( N_{a}\right)
}\rightarrow\upsilon^{a}}\;,\,\;H^{\upsilon}=\sum_{\underline{a}}\,\sum
_{s=1}^{N_{\underline{a}}-1}p_{\underline{a}}^{s}x_{s+1}^{\underline{a}}+p_{%
\underline{a}}^{N_{\underline{a}}}\upsilon^{\underline{a}}-L^{\upsilon }\,. 
\label{27}
\end{align}
The variables $\,p_{\underline{a}}^{s}\,,s=1,...,N_{\underline{a}}\,,$
should be treated as conjugate momenta to the coordinates $x_{s}^{\underline{%
a}}.$\ The corresponding Euler-Lagrange EM read:%
\begin{align}
\left. 
\begin{array}{c}
\frac{\delta S^{\upsilon}}{\delta p_{\underline{a}}^{s}}=\dot{x}_{s}^{%
\underline{a}}-x_{s+1}^{\underline{a}}=0,\;s=1,...,N_{\underline{a}}-1\,, \\ 
\frac{\delta S^{\upsilon}}{\delta p_{\underline{a}}^{N_{\underline{a}}}}=%
\dot{x}_{N_{\underline{a}}}^{\underline{a}}-\upsilon^{\underline{a}}=0%
\end{array}
\right\} & \Longrightarrow\left\{ 
\begin{array}{c}
\dot{x}_{s}^{\underline{a}}=\left\{ x_{s}^{\underline{a}}\,,H^{\upsilon
}\right\} \,, \\ 
s=1,...,N_{\underline{a}}%
\end{array}
\right. ;  \label{28} \\
\left. 
\begin{array}{c}
\frac{\delta S^{\upsilon}}{\delta x_{1}^{\underline{a}}}=\frac{\partial
L^{\upsilon}}{\partial x_{1}^{\underline{a}}}-\dot{p}_{1}^{\underline{a}%
}=0\,, \\ 
\frac{\delta S^{\upsilon}}{\delta x_{s}^{\underline{a}}}=\frac{\partial
L^{\upsilon}}{\partial x_{s}^{\underline{a}}}-\,p_{\underline{a}}^{s-1}-\dot{%
p}_{\underline{a}}^{s}=0,\;s=2,...,N_{\underline{a}}%
\end{array}
\right\} & \Longrightarrow\left\{ 
\begin{array}{c}
\dot{p}_{\underline{a}}^{s}=\left\{ p_{\underline{a}}^{s}\,,\,H^{\upsilon
}\right\} , \\ 
s=1,...,N_{\underline{a}}%
\end{array}
\right. ;  \label{29} \\
\left. 
\begin{array}{c}
\frac{\delta S^{\upsilon}}{\delta\upsilon^{\bar{a}}}=\frac{\partial
L^{\upsilon}}{\partial\upsilon^{\bar{a}}}=0\,, \\ 
\frac{\delta S^{\upsilon}}{\delta\upsilon^{\underline{a}}}=\frac{\partial
L^{\upsilon}}{\partial\upsilon^{\underline{a}}}-p_{\underline{a}}^{N_{%
\underline{a}}}=0%
\end{array}
\right\} & \Longrightarrow\frac{\partial H^{\upsilon}}{\partial\upsilon ^{a}}%
=0\,.   \label{30}
\end{align}
It is easy to verify that after eliminating $p_{\underline{a}%
}^{s}\,,\;s=1,...,N_{\underline{a}},\;x_{s}^{\underline{a}}\,,\;s=2,...,N_{%
\underline{a}}\,,\;\upsilon^{\underline{a}}$ (these variables are auxiliary)
from the action $S^{\upsilon}$ and from Eqs. (\ref{28})-(\ref{30}) we arrive
at the initial action $S$ and at the Euler-Lagrange EM (\ref{20}). The phase
space is formed by the variables $x_{s}^{\underline{a}}\,,\,p_{\underline{a}%
}^{s}$ only, and the extended phase space is formed by the variables $%
\;x_{s}^{\underline{a}}\,,\,p_{\underline{a}}^{s};$ $\upsilon^{a}\,.$ By
effecting the Hamiltonization, we\ must try to eliminate the velocities $%
\upsilon^{a}$\ from the action $\;S^{\upsilon}$.

Consider the nonsingular case according to the definition (\ref{21}):

\begin{equation}
\tilde{M}=\det\left| \left| \frac{\partial^{2}L^{\upsilon}}{\partial
\upsilon^{a}\partial\upsilon^{b}}\right| \right| =\det\left| \left| \frac{%
\partial^{2}H^{\upsilon}}{\partial\upsilon^{a}\partial\upsilon^{b}}\right|
\right| \neq0\,.   \label{31}
\end{equation}
Thus, Eqs. (\ref{30}) can be solved with respect to all the velocities $%
\upsilon,$ such that these velocities can be expressed in the form 
\begin{equation*}
\upsilon=\bar{\upsilon}\left( x,p_{\underline{a}}^{N_{\underline{a}}}\right)
\,,\;\left. \frac{\partial H^{\upsilon}}{\partial\upsilon^{a}}\right|
_{\upsilon=\bar{\upsilon}}\equiv0\,. 
\end{equation*}
Now we can eliminate the variables $\upsilon$\ from the action $%
\;S^{\upsilon },\;$since they are auxiliary variables (see Appendix). Thus,\
the action $S^{\upsilon}$ and Eqs. (\ref{28})-(\ref{29}) are transformed
into the ordinary Hamilton action $S_{\mathrm{H}}$ and Hamilton EM for the
unconstrained phase-space variables $x_{s}^{a},p_{a}^{s}$:

\begin{align}
& S_{\mathrm{H}}=\int\left( \sum_{\underline{a}}\sum_{s=1}^{N_{\underline {a}%
}}p_{\underline{a}}^{s}\dot{x}_{s}^{\underline{a}}-H\right) dt\,,\;H=\left.
H^{\upsilon}\right| _{\upsilon=\bar{\upsilon}}\,,  \notag \\
& \dot{x}_{s}^{\underline{a}}=\left\{ x_{s}^{\underline{a}}\,,H\right\}
\,,\,\,\,\dot{p}_{\underline{a}}^{s}=\left\{ p_{\underline{a}%
}^{s}\,,\,H\right\} ,\,\ s=1,...,N_{\underline{a}}\,.   \label{33}
\end{align}
One can see that the Hamiltonian $H$ is the energy written in the
phase-space variables. Eqs. (\ref{33}) are solved with respect to the
highest (here first order) time derivatives and therefore have a unique
solution whenever $2\sum_{\underline{a}}N_{\underline{a}}=2\sum_{a}N_{a}\;$%
initial data are given. Since these Hamilton EM are equivalent to the
Euler-Lagrange EM (\ref{20}), we can say that for nonsingular (according to
the new definition (\ref{21})) theories, the EM have a unique solution
whenever $2\sum_{a}N_{a}$ initial data are given! Of course this fact could
be established directly in the Lagrangian formulation.

Consider now the singular case\footnote{%
We suppose that the generalized Hessian matrix has a constant rank in a
vicinity of the consideration point.}

\begin{align}
& \tilde{M}=\det\left| \left| \frac{\partial^{2}L^{\upsilon}}{%
\partial\upsilon^{a}\partial\upsilon^{b}}\right| \right| =\det\left| \left| 
\frac{\partial^{2}H^{\upsilon}}{\partial\upsilon^{a}\partial \upsilon^{b}}%
\right| \right| =0\,,  \notag \\
& \mathrm{rank}\left| \left| \frac{\partial^{2}L^{\upsilon}}{\partial
\upsilon^{a}\partial\upsilon^{b}}\right| \right| =\mathrm{rank}\left| \left| 
\frac{\partial^{2}H^{\upsilon}}{\partial\upsilon^{a}\partial \upsilon^{b}}%
\right| \right| =R,\;n-R>0\,.   \label{32}
\end{align}
In the singular case, the equations (\ref{30}) do not allow to express all
the velocities through $x_{s}^{\underline{a}},p_{\underline{a}}^{N_{%
\underline{a}}}$, so part of the velocities appear as the Lagrange
multipliers to primary constraints. Let us see how it works. Without loss of
generality, we now number the coordinates in such a way that in the
generalized Hessian matrix $\tilde{M}$ (or in the matrix $\tilde{M}%
^{\upsilon}$) the nonzero minor of maximum size $R$ is placed in the top
left corner. This is always possible, because in a symmetric matrix a
principle nonzero minor of maximum size exists. Then, the velocities $%
\upsilon$ are divided into two groups:%
\begin{align*}
& \upsilon^{i}\,=V^{i}\,,\,\,\,\,i=1\,,...\,,R\,,\;\upsilon^{R+\varkappa
}=\lambda^{\varkappa}\,,\;\;\varkappa=1,...\,,n-R\,, \\
& \det\left\| \frac{\partial^{2\,}L^{\upsilon}}{\partial^{2}V^{i}\partial
V^{j}}\right\| \neq0\,.
\end{align*}
The indices $i\;$and $\varkappa\,$are divided as $i=(\underline{i},\bar {%
\imath})$\ and $\varkappa=(\underline{\varkappa},\bar{\varkappa})\,,$ such
that $\underline{a}=\left( \underline{i},\underline{\varkappa}\right) $,$\;%
\bar{a}=\left( \bar{\imath},\bar{\varkappa}\right) $. Owing to the
fulfillment of the conditions (\ref{r1}), all the velocities $V$ can be
expressed with the help of the equations\footnote{%
Throughout this section we use a notation of the type $\left. F\left(
\upsilon\right) \right| _{V=\bar{V}}=\overline{F\left( \upsilon\right) }\;.$}
\begin{equation}
\frac{\partial H^{\upsilon}}{\partial V^{i}}=0   \label{4.17}
\end{equation}
via the variables $q^{\underline{a}},p_{\underline{a}};\lambda,$

\begin{equation}
V^{i}=\bar{V}^{i}\left( x,p_{\underline{i}}^{N_{\underline{i}%
}};\lambda\right) \,,\;\overline{\frac{\partial H^{\upsilon}}{\partial V}}%
\equiv0\,.   \label{4.18}
\end{equation}
We shall call the velocities $V$ primarily expressible velocities,\emph{\ }%
and the remaining velocities $\lambda$ will be called primarily
unexpressible velocities.\emph{\ }Substituting the primarily expressible
velocities into the remaining equations

\begin{equation}
\frac{\partial H^{\upsilon}}{\partial\lambda^{\varkappa}}=0\,,   \label{4.19}
\end{equation}
we arrive at the relations (primary constraints):%
\begin{equation}
\Phi_{\varkappa}^{(1)}\left( x,p_{\underline{a}}^{N_{\underline{a}}}\right) =%
\overline{\frac{\partial H^{\upsilon}}{\partial\lambda^{\varkappa}}}%
\,=0\,,\;\varkappa=1,...\,,n-R\,.   \label{4.20}
\end{equation}
The left-hand sides of the relations (\ref{4.20}) contain no primarily
unexpressible velocities. Indeed, if at least one of the velocities $\lambda$
were contained in (\ref{4.20}), we would be able to express it in terms of
the rest of the variables, and this contradicts our supposition about the
ranks. The latter qualitative argument can be confirmed by a strict
consideration based on our supposition about ranks. There are two types of
primary constraints (\ref{4.20}):

\begin{align}
& \Phi_{\underline{\varkappa}}^{(1)}\,=p_{\underline{\varkappa}}^{N_{%
\underline{\varkappa}}}+f_{\underline{\varkappa}}\left( x,p_{\underline {i}%
}^{N_{\underline{i}}}\right) =0\,,  \label{4.21} \\
& \Phi_{\bar{\varkappa}}^{(1)}\,=f_{\bar{\varkappa}}\left( x,p_{\underline {i%
}}^{N_{\underline{i}}}\right) =0\,,  \label{r2} \\
& f_{\varkappa}\left( x,p_{\underline{i}}^{N_{\underline{i}}}\right) =-%
\overline{\frac{\partial L^{\upsilon}}{\partial\lambda^{\varkappa}}}\,. 
\label{r4}
\end{align}
All the constraints $\Phi_{\underline{\varkappa}}^{(1)}$ are independent
among themselves and of the constraints $\Phi_{\bar{\varkappa}}^{(1)}.$ The
latter may be dependent.

The velocities $V$ are auxiliary variables and can be excluded from the
action. Let us substitute $V=\bar{V}$ into the action (\ref{27}). First we
write the Hamiltonian $H^{\upsilon}$ in the following form:%
\begin{align}
& H^{\upsilon}=\sum_{\underline{a}}\left[ \sum_{s=1}^{N_{\underline{a}}-1}p_{%
\underline{a}}^{s}x_{s+1}^{\underline{a}}+p_{\underline{a}}^{N_{\underline{a}%
}}\upsilon^{\underline{a}}\right] -L^{\upsilon}\equiv\frac{\partial
L^{\upsilon}}{\partial\upsilon^{\underline{a}}}\upsilon^{\underline{a}}+%
\frac{\partial L^{\upsilon}}{\partial\upsilon ^{\bar{a}}}\upsilon^{\bar{a}} 
\notag \\
& +\sum_{\underline{a}}\sum_{s=1}^{N_{\underline{a}}-1}p_{\underline{a}%
}^{s}x_{s+1}^{\underline{a}}-L^{\upsilon}+\upsilon^{\underline{a}}\left( p_{%
\underline{a}}^{N_{\underline{a}}}-\frac{\partial L^{\upsilon}}{%
\partial\upsilon^{\underline{a}}}\right) -\upsilon^{\bar{a}}\frac{\partial
L^{\upsilon}}{\partial\upsilon^{\bar{a}}}  \notag \\
& \equiv\left( \frac{\partial L^{\upsilon}}{\partial\upsilon^{a}}%
\upsilon^{a}+\sum_{\underline{a}}\sum_{s=1}^{N_{\underline{a}}-1}p_{%
\underline{a}}^{s}x_{s+1}^{\underline{a}}-L^{\upsilon}\right) +\upsilon^{a}%
\frac{\partial H^{\upsilon}}{\partial\upsilon^{a}}\equiv \mathcal{E}%
^{\upsilon}+\upsilon^{a}\frac{\partial H^{\upsilon}}{\partial \upsilon^{a}}%
\,,  \label{r5} \\
& \mathcal{E}^{\upsilon}=\left( \frac{\partial L^{\upsilon}}{\partial
\upsilon^{a}}\upsilon^{a}+\sum_{\underline{a}}\sum_{s=1}^{N_{\underline{a}%
}-1}p_{\underline{a}}^{s}x_{s+1}^{\underline{a}}-L^{\upsilon}\right) \,. 
\label{r6}
\end{align}
One can see that $\mathcal{E}^{\upsilon}$ coincides (on the equations of
motion (\ref{28}-\ref{30})) with the usual the Lagrangian energy

\begin{equation}
\mathcal{E}=\sum_{\underline{a}}\sum_{l=1}^{N_{\underline{a}}}q^{\underline {%
a}\left( l\right) }\sum_{s=l}^{N_{\underline{a}}}\left( -1\right) ^{^{s-l}}%
\frac{d^{s-l}}{dt^{s-l}}\frac{\partial L}{\partial q^{\underline {a}\left(
s\right) }}-L\,.   \label{r10}
\end{equation}
Then we substitute $V$ as $V=\bar{V}$ into the Hamiltonian $H^{\upsilon}.$
Thus we obtain the total Hamiltonian $H^{\left( 1\right) }$: 
\begin{equation}
H^{\left( 1\right) }=\overline{H^{\upsilon}}\,=H+\lambda^{\varkappa}\Phi_{%
\varkappa}^{\left( 1\right) }\,,\;\;H=\overline{\mathcal{E}^{\upsilon }}\,. 
\label{r7}
\end{equation}
Let us study the structure of the Hamiltonian $H^{\left( 1\right) }.$ We
compare the relation%
\begin{equation*}
\frac{\partial H^{\left( 1\right) }}{\partial\lambda^{\varkappa}}=\frac{%
\partial H}{\partial\lambda^{\varkappa}}+\Phi_{\varkappa}^{\left( 1\right)
}\,\, 
\end{equation*}
with the identity%
\begin{equation*}
\frac{\partial H^{\left( 1\right) }}{\partial\lambda^{\varkappa}}\equiv%
\overline{\frac{\partial H^{\upsilon}}{\partial\lambda^{\varkappa}}}+%
\overline{\frac{\partial H^{\upsilon}}{\partial V^{i}}}\frac{\partial\bar
{V%
}^{i}}{\partial\lambda^{\varkappa}}\equiv\Phi_{\varkappa}^{\left( 1\right)
}\,\, 
\end{equation*}
to conclude that $H=H\left( x,p\right) \,.$

Thus in the singular case, after the generalized Hamiltonization procedure,
we arrive at $the$ Hamiltonian theory with primary constraints whose action
is 
\begin{equation}
S^{\left( 1\right) }=\int\left[ \sum_{\underline{a}}\sum_{s=1}^{N_{%
\underline{a}}}p_{\underline{a}}^{s}\dot{x}_{s}^{\underline{a}}-H^{\left(
1\right) }\right] dt\,,\;H^{\left( 1\right) }\,=H+\lambda
^{\varkappa}\Phi_{\varkappa}^{\left( 1\right) }\,.   \label{r9}
\end{equation}
Further Hamiltonization is proceeded\ according to the usual Dirac procedure.

One ought to stress that the generalized Hamiltonization procedure and the
usual Hamiltonization procedure lead to formally different (but equivalent)
Hamiltonian theories with primary constraints. After the generalized
Hamiltonization procedure, some of the degenerate coordinates may appear as
Lagrange multipliers in the total Hamiltonian. The corresponding primary
constraints may be dependent. However, if we denote via $\Phi_{\bar{%
\varkappa }^{\prime}}^{(1)}$ the independent constraints (\ref{r2}), then%
\begin{equation*}
\Phi_{\bar{\varkappa}}^{\left( 1\right) }=S_{\bar{\varkappa}}^{\bar {%
\varkappa}^{\prime}}\Phi_{\bar{\varkappa}^{\prime}}^{(1)}, 
\end{equation*}
where $S$ is some matrix. Thus, $\Phi_{\underline{\varkappa}}^{(1)},\Phi _{%
\bar{\varkappa}^{\prime}}^{(1)}$ is a complete set of independent primary
constraints, and the total Hamiltonian takes the form%
\begin{equation*}
H^{\left( 1\right) }=H+\bar{\lambda}^{\varkappa^{\prime}}\Phi_{\varkappa
^{\prime}}^{(1)},\;\varkappa^{\prime}=(\underline{\varkappa},\bar{\varkappa }%
^{\prime})\,, 
\end{equation*}
where $\bar{\lambda}\;$are some new Lagrange multipliers. Then further
Hamiltonization is proceeded\ according to the usual Dirac procedure.

We see that by performing the Hamiltonization procedure one does not need,
in principle, to introduce the momenta conjugate to the degenerate
coordinates. The singularity definition (\ref{21}) differs from (\ref{7c})
for theories with degenerate coordinates and seems\ more reasonable in such
cases. As was already remarked, considering the aforementioned instructive
example, besides the natural consistency of the generalized Hamiltonization
procedure, the generalized-Hessian criterion allows one to conclude
immediately that the theory is not a gauge theory when $\tilde{M}\neq0$.
Examining only the conventional Hessian, we cannot come to such a conclusion
without additional analysis of the constraint structure.

If we select some $N_{a}>\bar{N}_{a},$ a wider extended phase space is
needed for the Hamiltonization. However, one can demonstrate, similarly to %
\cite{GitTyL85,GitTy90}, that the resulting Hamiltonian theory will be
equivalent to the one with orders $\{N_{a}=\bar{N}_{a}\}$.

\section{Examples and discussion}

1) Consider the theory with degenerate coordinate $u$ and Lagrange function
of the form 
\begin{equation*}
L=\dot{x}u-V\left( x,u\right) . 
\end{equation*}
Selecting $N_{x}=N_{u}=1,$ we must introduce the momenta $p$ to $x$ and $%
p^{\prime}$ to $u$. In the first-order formalism, the phase space is formed
by the pairs $x,p;\,u,p^{\prime},$ and the extended phase space is formed by
the variables $x,p;\,u,p^{\prime};\upsilon,\upsilon^{\prime}.$ The
first-order formalism action reads:

\begin{equation}
S^{\upsilon\upsilon^{\prime}}=\int\left[ p\dot{x}+p^{\prime}\dot {u}%
-H^{\upsilon\upsilon^{\prime}}\right] dt\,,\,\;H^{\upsilon\upsilon
^{\prime}}=p\upsilon+p^{\prime}\upsilon^{\prime}-\upsilon u+V\left(
x,u\right) \,.   \label{e1}
\end{equation}
From the equations $\partial H^{\upsilon\upsilon^{\prime}}/\partial
\upsilon=\partial H^{\upsilon\upsilon^{\prime}}/\partial\upsilon^{\prime}=0$
we find two primary constraints $\Phi_{i}^{\left( 1\right) }=0,\;i=1,2$, 
\begin{equation}
\Phi_{1}^{\left( 1\right) }=p-u,\;\;\;\Phi_{2}^{\left( 1\right)
}=p^{\prime}.   \label{e2}
\end{equation}
They are second-class,$\;\left\{ \Phi_{1}^{\left( 1\right) },\Phi
_{2}^{\left( 1\right) }\right\} =-1;$ no more constraints appear. The total
Hamiltonian reads: 
\begin{equation*}
H^{\left( 1\right) }=V\left( x,y\right) +\lambda^{i}\Phi_{i}^{\left(
1\right) },\;\lambda^{1}=\upsilon,\;\lambda^{2}=\upsilon^{\prime}\,. 
\end{equation*}
One can use the constraints \ref{e2} to exclude the variables $p^{\prime}$
and $u$ from the action (\ref{e1}). Thus, we get: 
\begin{equation}
S_{\mathrm{H}}=\int\left[ p\dot{x}-V\left( x,p\right) \right] dt\,\,. 
\label{e2a}
\end{equation}
For the unconstrained variables $x$ and $p$ we obtain ordinary Hamilton
equations with the Hamiltonian\ $H=V\left( x,p\right) $.

If we choose $N_{x}=1$ and $N_{u}=0$ then we have to introduce the momentum $%
p$ to $x$ only. In the first-order formalism, the phase space is formed by
the pair $x,p,$ and the extended phase space is formed by the variables $%
x,p;\upsilon.$ The first-order formalism action reads:

\begin{equation}
S^{\upsilon}=\int\left[ p\dot{x}-H^{\upsilon}\right] dt\,,\,\;H^{\upsilon
}=p\upsilon-\upsilon u+V\left( x,u\right) \,.   \label{e3}
\end{equation}
The generalized Hessian $\tilde{M}$ equals $-1$, thus, the theory is
nonsingular in the new definition. Both variables $\upsilon,u$ are
auxiliary. The equations $\partial
H^{\upsilon}/\partial\upsilon=0,\;\partial H^{\upsilon}/\partial u=0$ allow
one to find these variables as: $u=p,\;\upsilon=\partial V\left( x,p\right)
/\partial p$ and exclude them from (\ref{e3}) to get immediately the action (%
\ref{e2a}).Thus, we see that in the framework of the generalized
Hamiltonization procedure we arrive to the result in a more simple way.

2. Consider the theory with degenerate coordinate $u$ and Lagrange function
of the form

\begin{equation*}
L=\dot{x}u+\dot{x}^{3}. 
\end{equation*}
The Euler-Lagrange EM give $\dot{u}=\dot{x}=0\,.$ Here not only the Hessian
is zero, but also the Hessian matrix

\begin{equation*}
\left| \left| 
\begin{array}{cc}
\frac{\partial^{2}L}{\partial\dot{x}\partial\dot{x}} & \frac{\partial^{2}L}{%
\partial\dot{x}\partial\dot{u}} \\ 
\frac{\partial^{2}L}{\partial\dot{u}\partial\dot{x}} & \frac{\partial^{2}L}{%
\partial\dot{u}\partial\dot{u}}%
\end{array}
\right| \right| =\left| \left| 
\begin{array}{ll}
6\dot{x} & 0 \\ 
0 & 0%
\end{array}
\right| \right| \, 
\end{equation*}
does not have a constant rank in the vicinity of the zero point $x=\dot
{x}%
=u=\dot{u}=0$, such that we may have additional difficulties when using the
usual Hamiltonization procedure. In contrast to this, the generalized
Hessian $\tilde{M}$ equals $-1.$ Thus, the generalized Hamiltonization
procedure can be completed without such difficulties. This example gives
additional arguments in favour of the generalized Hamiltonization procedure
for theories with degenerate coordinates.

3. Consider the theory of a massive vector field $A^{\mu}$ . The theory is
described by the Proca action

\begin{align}
& S=\int\mathcal{L}dx\,,\;\mathcal{L=}-\frac{1}{4}F_{\mu\nu}F^{\mu\nu }+%
\frac{m^{2}}{2}A_{\mu}A^{\mu}  \notag \\
& \,=\frac{1}{2}\left( \dot{A}^{i}+\partial_{i}A^{0}\right) ^{2}-\frac{1}{4}%
F_{ik}F^{ik}+\frac{m^{2}}{2}A_{\mu}A^{\mu},\;F_{\mu\nu}=\partial_{\mu}A_{%
\nu}-\partial_{\nu}A_{\mu}\,.   \label{50}
\end{align}
In this case the velocity $\dot{A}^{0}$ does not enter the Lagrangian. Thus, 
$A^{0}$ is a degenerate variable. Below we compare the conventional and the
generalized Hamiltonization procedures.

a) Selecting all $N_{\mu}=1,$ we consider the conventional Hamiltonization.
Here we see that the Hessian is zero, and the theory is singular in the
conventional definition. We introduce momenta $p_{\mu}$ to all the
coordinates $A^{\mu}$\thinspace. The action $S^{\upsilon}$ of the
first-order formalism reads

\begin{align}
& S^{\upsilon}=\int\left[ p_{\mu}\dot{A}^{\mu}-\mathcal{H}^{\upsilon }\right]
dx\,,  \notag \\
& \mathcal{H}^{\upsilon}=p_{\mu}\upsilon^{\mu}-\frac{1}{2}\left[ \left(
\upsilon^{i}+\partial_{i}A_{0}\right) ^{2}-\frac{1}{2}F_{ik}F^{ik}+\frac{%
m^{2}}{2}A_{\mu}A^{\mu}\right] \,.   \label{51a}
\end{align}
Thus, $A^{\mu},p_{\mu}$ form the phase space and $A^{\mu},$ $%
p_{\mu};\upsilon^{\mu}$ form the extended phase space. The equations 
\begin{equation}
\delta S^{\upsilon}/\delta\upsilon^{\mu}=0\Leftrightarrow\left\{ 
\begin{array}{l}
p_{0}=\frac{\partial\mathcal{L}}{\partial\upsilon^{0}}=0\, \\ 
p_{i}=\frac{\partial\mathcal{L}}{\partial\upsilon^{i}}=\upsilon^{i}+%
\partial_{i}A^{0}\,,%
\end{array}
\right.   \label{51b}
\end{equation}
do not allow one to express the velocity $\upsilon^{0}$ via the other
variables, there is a primary constraint $\Phi^{(1)}=p_{0}=0$. At the same
time the velocities $\upsilon^{i}$ are auxiliary variables (see Appendix),
so they can be expressed via the other variables by aid of equations (\ref%
{51b}), $\upsilon^{i}=\bar{\upsilon}^{i}\left( p,A\right)
=p_{i}-\partial_{i}A^{0},$ and substituted into (\ref{51a}) to get the
reduced equivalent action

\begin{align}
& S^{\left( 1\right) }=\int\left[ p_{\mu}\dot{A}^{\mu}-\mathcal{H}^{\left(
1\right) }\right] dx\,,\;\mathcal{H}^{\left( 1\right) }=\left. \mathcal{H}%
^{\upsilon}\right| _{\upsilon^{i}=\bar{\upsilon}^{i}\left( p,A\right) }=%
\mathcal{H}+\lambda\Phi^{(1)},  \notag \\
& \mathcal{H}=\left. \left( \frac{\partial\mathcal{L}}{\partial\upsilon ^{i}}%
\upsilon^{i}-\mathcal{L}\right) \right| _{\upsilon^{i}=\bar{\upsilon }%
^{i}\left( p,A\right) }=\frac{1}{2}p_{i}^{2}-p_{i}\partial_{i}A^{0}+\frac{1}{%
4}F_{ik}F^{ik}-\frac{m^{2}}{2}A_{\mu}A^{\mu}\,,\,   \label{51c}
\end{align}
where $\lambda=\upsilon^{0}.$ Applying the consistency condition to the
primary constraint, we find the secondary constraint $\Phi^{(2)}=\partial
_{i}p_{i}-m^{2}A^{0}=0\,.$ There are no further secondary constraints and $%
\Phi=(\Phi^{(1)},\Phi^{(2)})$ is the complete set of second-class
constraints. Moreover, this set of constraints $\Phi$ is of the special form %
\cite{GitTy90}, such that one can use these constraints to eliminate the
variables $A^{0}$ and $p_{0}$ from the action (\ref{51c}) to get reduced
equivalent Hamilton action

\begin{align}
& S_{\mathrm{H}}=\int\left[ p_{i}\dot{A}^{i}-\mathcal{\tilde{H}}\right]
dx\,,\;\;\mathcal{\tilde{H}}=\left. \mathcal{H}\right|
_{A^{0}=m^{-2}\partial_{i}p_{i}}  \label{51d} \\
& \mathcal{\,}=\frac{1}{2}p_{i}^{2}+\frac{1}{2m^{2}}\left(
\partial_{i}p_{i}\right) ^{2}+\frac{1}{4}F_{ik}F^{ik}+\frac{m^{2}}{2}%
A_{i}^{2}\,.
\end{align}
It follows from (\ref{51d}) that\ in the phase space $A^{i},p_{i}$ the
dynamics is governed according to ordinary Hamilton EM with the Hamiltonian
density $H=\int\mathcal{\tilde{H}}d\mathbf{x}$ and without any constraints.

b) In the generalized Hamiltonization procedure we select $N_{i}=1,\;N_{0}=0,
$ and introduce the velocities according to the general prescription (see
Sect.III) as $\upsilon^{0}=A^{0},\;\upsilon^{i}=\dot{A}^{i}\,.$ The theory
is not singular with respect to the new definition (\ref{21}), since the
matrix $\partial^{2}\mathcal{L}/\partial\upsilon^{\mu}\partial\upsilon^{\nu}$
is invertible. We introduce the momenta $p_{i}$ conjugate to the coordinates 
$A^{i}$ only. Thus, $A^{i},p_{i}$ form the phase space and $A^{i},$ $%
p_{i};\upsilon^{\mu}$ form the extended phase space. The action $S^{\upsilon
}$ in the first-order formalism reads

\begin{align}
& S^{\upsilon}=\int\left[ p_{i}\dot{A}^{i}-\mathcal{H}^{\upsilon}\right]
dx\,,  \notag \\
& \mathcal{H}^{\upsilon}=p_{i}\upsilon^{i}-\frac{1}{2}\left[ \left(
\upsilon^{i}+\partial_{i}\upsilon^{0}\right) ^{2}-\frac{1}{2}%
F_{ik}F^{ik}+m^{2}\left( \upsilon_{0}^{2}-A_{i}^{2}\right) \right] \,. 
\label{54}
\end{align}
Here the equations $\delta S^{\upsilon}/\delta\upsilon^{\mu}=0$ allow one to
express all the velocities via the momenta (all the velocities are auxiliary
variables) as $\upsilon=\bar{\upsilon}\left( p\right) ,$

\begin{equation}
\upsilon^{0}=m^{-2}\partial_{i}p_{i}\,,\;\upsilon^{i}=\left( \delta
_{ij}-m^{-2}\partial_{i}\partial_{j}\right) p_{j\,}.   \label{55}
\end{equation}
Substituting the expressions (\ref{55}) into (\ref{54}) we obtain
immediately the reduced equivalent Hamilton action (\ref{51d}). Remark that 
\begin{equation*}
\mathcal{\tilde{H}}=\left. \left( \frac{\partial\mathcal{L}}{\partial
\upsilon^{\mu}}\upsilon^{\mu}-\mathcal{L}\right) \right| _{\upsilon =\bar{%
\upsilon}\left( p\right) }. 
\end{equation*}

4. Consider electrodynamics. The theory is described by the Maxwell action,

\begin{equation}
S=\int\mathcal{L}dx\,,\;\mathcal{L=}-\frac{1}{4}F_{\mu\nu}F^{\mu\nu}=\frac{1%
}{2}\left( \dot{A}^{i}+\partial_{i}A^{0}\right) ^{2}-\frac{1}{4}%
F_{ik}F^{ik}\,.   \label{53}
\end{equation}
Here both the Hessian and the generalized Hessian are zero, thus the theory
is singular in both definitions. As before $A^{0}$ is a degenerate variable.
Let us compare the conventional and the generalized Hamiltonization
procedures.

a) First consider the conventional Hamiltonization procedure, selecting all $%
N_{\mu}=1$. The theory is singular. Here we have to introduce momenta $%
p_{\mu}$ conjugate to all the coordinates $A^{\mu}$\thinspace. The action $%
S^{\upsilon}$ of the first-order formalism reads

\begin{equation}
S^{\upsilon}=\int\left[ p_{\mu}\dot{A}^{\mu}-\mathcal{H}^{\upsilon}\right]
dx\,,\;\mathcal{H}^{\upsilon}=p_{\mu}\upsilon^{\mu}-\frac{1}{2}\left[ \left(
\upsilon^{i}+\partial_{i}A_{0}\right) ^{2}-\frac{1}{2}F_{ik}F^{ik}\right]
\,.   \label{59}
\end{equation}
Thus, $A^{\mu},p_{\mu}$ form the phase space and $A^{\mu},$ $%
p_{\mu};\upsilon^{\mu}$ form the extended phase space. As in the Proca case,
we have here a primary constraint $\Phi^{(1)}=p_{0}=0,$ primarily
expressible velocities $\upsilon^{i}=\bar{\upsilon}^{i}\left( p,A\right)
=p_{i}-\partial_{i}A^{0}.\;$The reduced equivalent action reads

\begin{align}
& S^{\left( 1\right) }=\int\left[ p_{\mu}\dot{A}^{\mu}-\mathcal{H}^{\left(
1\right) }\right] dx\,,\;\mathcal{H}^{\left( 1\right) }=\left. \mathcal{H}%
^{\upsilon}\right| _{\upsilon^{i}=\bar{\upsilon}^{i}\left( p,A\right) }=%
\mathcal{H}+\lambda\Phi^{(1)},  \notag \\
& \mathcal{H}=\left. \left( \frac{\partial\mathcal{L}}{\partial\upsilon ^{i}}%
\upsilon^{i}-\mathcal{L}\right) \right| _{\upsilon^{i}=\bar{\upsilon }%
^{i}\left( p,A\right) }=\frac{1}{2}p_{i}^{2}-p_{i}\partial_{i}A^{0}+\frac{1}{%
4}F_{ik}F^{ik},\,   \label{60}
\end{align}
where $\lambda=\upsilon^{0}.$ Applying the consistency condition to the
primary constraint, we find the secondary constraint $\Phi^{(2)}=\partial
_{i}p_{i}=0,\,\ $and no further constraints arise. One can see that $%
\Phi=(\Phi^{(1)},\Phi^{(2)})=0$ is a set of first-class constraints, such
that in this case $\lambda$ is undetermined. It is a gauge theory. Any rigid
gauge needs two additional gauge conditions (in particular, to fix $\lambda$%
), and one can see that $A^{0}$ is not a physical variable \cite{GitTy90}.

b) In the generalized Hamiltonization procedure, we select $%
N_{i}=1,\;N_{0}=0,$ and introduce the velocities according to the general
prescription (see Sect.III) as $\upsilon^{0}=A^{0},\;\upsilon^{i}=\dot{A}%
^{i}\,.$ The theory is singular with respect to the new definition (\ref{21}%
) as well, since the matrix $\partial^{2}\mathcal{L}/\partial\upsilon^{\mu
}\partial\upsilon^{\nu}$ is not invertible. We do not introduce momenta
conjugate to $A^{0}.$ The action of the first-order formalism $S^{\upsilon}$
reads

\begin{equation}
S^{\upsilon}=\int\left[ p_{i}\dot{A}^{i}-\mathcal{H}^{\upsilon}\right]
dx\,,\;\mathcal{H}^{\upsilon}=p_{i}\upsilon^{i}-\frac{1}{2}\left(
\upsilon^{i}+\partial_{i}\upsilon^{0}\right) ^{2}+\frac{1}{4}F_{ik}F^{ik}\,. 
\label{56}
\end{equation}
Here the equations $\delta S^{\upsilon}/\delta\upsilon^{\mu}=0\;$give $%
\delta
S^{\upsilon}/\delta\upsilon^{i}=0\Longrightarrow\upsilon^{i}=p_{i}-%
\partial_{i}\upsilon^{0}$ and the primary constraint $\Phi^{\left( 1\right)
}=\partial_{i}p_{i}=0\,.\;$Only the velocities $\upsilon^{i}$ are auxiliary
variables. They can be eliminated from the action using the corresponding
equations. Therefore, we are left with the reduced equivalent Hamilton action

\begin{align}
& S^{\left( 1\right) }=\int\left[ p_{i}\dot{A}^{i}-\mathcal{H}^{\left(
1\right) }\right] dx\,,\;\;\mathcal{H}^{\left( 1\right) }=\left. \mathcal{H}%
^{\upsilon}\right| _{\upsilon^{i}=p_{i}-\partial_{i}\upsilon^{0}}=\mathcal{H}%
+\lambda\Phi^{\left( 1\right) }\,,  \notag \\
& \mathcal{H}=\left. \left( \frac{\partial\mathcal{L}}{\partial
\upsilon^{\mu}}\upsilon^{\mu}-\mathcal{L}\right) \right| _{\upsilon
^{i}=p_{i}-\partial_{i}\upsilon^{0}}=\frac{1}{2}p_{i}^{2}+\frac{1}{4}%
F_{ik}F^{ik}\,,\;\;\lambda=A^{0}\,.   \label{58}
\end{align}
We see that in the generalized Hamiltonization procedure, the variable $A^{0}
$ has naturally the status of a Lagrange multiplier to the primary
constraint. The action (\ref{58}) is physically equivalent to (\ref{60}).
That can be seen, for example, by excluding the auxiliary variables $%
\;\lambda,\;p_{0}\;$from\ the action\ (\ref{60}).

\begin{acknowledgement}
Gitman is grateful to the foundations FAPESP, CNPq, DAAD for support as well
as to the Lebedev Physics Institute (Moscow) and to the Institute of
Theoretical Physics (University of Leipzig) for hospitality; Tyutin thanks
INTAS 00-00262, RFBR 02-02-16944, 00-15-96566 for partial support.
\end{acknowledgement}

\section*{Appendix}

Consider a classical system described by a set of generalized coordinates\ $%
q\equiv\{q^{a};a=1,2,...,n\}$ and by the action $S\left[ q\right] =\int
Ldt\,.$ Sometimes the EM allow one to uniquely express a part of the
variables, which we denote by $y$ , via the rest of the variables, which we
denote by $x$ , such that $q^{a}=\left( y^{i},x^{\mu}\right) $. In such a
case, one can try to eliminate the variables $y$ from the initial action and
ask whether the initial and the reduced theories are equivalent. In what
follows, we consider the case in which a positive answer is possible, see in
this regard \cite{Superflu,BorTy98}.

Suppose an action $S\left[ q\right] =S[y,x]$ is given such that the EM $%
\delta S/\delta y=0$ allow us to express uniquely the variables $y$ as local
functions of the variables $x,$ namely:

\begin{equation}
\frac{\delta S\left[ y,x\right] }{\delta y}=0\Longleftrightarrow y=\bar {y}%
(t,x^{\left( l\right) }).   \label{1}
\end{equation}
Consider the action $\bar{S}[x]\equiv\,S[\bar{y},x],$ which we call the
reduced action and compare the EM corresponding to both actions. Consider
the variation $\delta\bar{S}\,\ $under arbitrary inner variations $\delta x$
such that any surface terms vanish \footnote{%
To derive the equations of motion it is enough to only consider such inner
variations.},

\begin{equation}
\delta\bar{S}[x]=\int\left( \left. \frac{\delta S\left[ y,x\right] }{\delta y%
}\right| _{y=\bar{y}}\delta\bar{y}+\left. \frac{\delta S\left[ y,x\right] }{%
\delta x}\right| _{y=\bar{y}}\delta x\right) dt=\int \frac{\delta\bar{S}%
\left[ x\right] }{\delta x}\delta xdt\,.   \label{2}
\end{equation}
In virtue of (\ref{1}), the EM for the reduced theory read:

\begin{equation}
\frac{\delta\bar{S}\left[ x\right] }{\delta x}=\left. \frac{\delta S\left[
y,x\right] }{\delta x}\right| _{y=\bar{y}}=0\,.   \label{3}
\end{equation}

On the other hand, the EM for the initial theory are:

\begin{equation*}
\frac{\delta S\left[ y,x\right] }{\delta y}=0\Longleftrightarrow y=\bar
{y}%
(x,\dot{x},...)\,;\;\;\frac{\delta S\left[ y,x\right] }{\delta x}=0\,. 
\end{equation*}
They are reduced to (\ref{3}) in the sector of the $x$-variables $.$ Thus,
the initial action $S$ and the reduced action $\bar{S}$ lead to the same EM
for the variables $x$ . For this reason, we can treat the $y$-variables as\
dependent and call them auxiliary variables. Thus, the auxiliary variables
can be eliminated with the help of the EM derived from the action. The
initial theory and the reduced one are equivalent. One ought to stress that
this equivalence is a consequence of supposition (\ref{1}), that is, it is
important that the $y$-variables be expressed as functionals of $x$ by means
of the equations $\delta S/\delta y=0$ only. If for this purpose some of the
equations $\delta S/\delta x=0$ are used, then the above equivalence may be
untrue. Of course, solutions of the reduced\ theory, together with\emph{\ }%
the relation $y=\bar{y}$,$\;$contain\emph{\ }all\emph{\ }the\emph{\ }%
solutions of the initial theory\emph{\ }as it is easily seen\ from\ Eq. (\ref%
{2}).\ However, the reduced theory may have additional solutions. To
illustrate this fact, we consider a Lagrange function of the form

\begin{equation}
L=\dot{x}^{2}/2+xy\,.   \label{5}
\end{equation}
The corresponding EM are $\delta S/\delta x=y-\ddot{x}=0,\quad\delta
S/\delta y=x=0\,.$ They have the unique solution $x=y=0.$ Now let us express
the variable $y$ via $x$ using the equation $\delta S/\delta x=0.$ The
reduced Lagrange function takes the form

\begin{equation}
\bar{L}=\dot{x}^{2}/2+x\ddot{x}\,.   \label{6}
\end{equation}
We see that the EM $\delta\bar{S}/\delta x=\ddot{x}=0$ of the reduced theory
together with $y=\ddot{x}$ have additional solutions in comparison with the
initial theory. If we use the equation $\delta S/\delta y=x=0$ for
eliminating $x$ from the Lagrange function, then the reduced theory (for the
variable $y$ with $\bar{L}=0)$ becomes a gauge theory, whereas the initial
theory was not.

\end{document}